\documentclass[accepted]{uai2026} % after acceptance, for a revised version; 
\usepackage[american]{babel}

\usepackage{amsmath,amssymb,amsfonts}
\usepackage{multirow}
\usepackage{tabularx}
\usepackage{array}
\usepackage{algpseudocode}
\usepackage{caption}
\usepackage{algorithmicx}
\usepackage{algorithm,algpseudocode}
\usepackage{longtable}
\usepackage{tabularray}
\UseTblrLibrary{booktabs}
\usepackage{cuted}
\usepackage{subcaption}

\providecommand{\qedsymbol}{$\square$}
\newenvironment{proof}[1][Proof]{\par\vspace{1ex}\noindent\textbf{#1.} \hspace{0.2em}}{\hfill\qedsymbol\vspace{1ex}\par}

\newtheorem{theorem}{Theorem}[section]
\newtheorem{lemma}[theorem]{Lemma}
\newtheorem{proposition}[theorem]{Proposition}

\newtheorem{definition}[theorem]{Definition}

\newtheorem{example}[theorem]{Example}

\usepackage{natbib} % has a nice set of citation styles and commands
\usepackage{mathtools} % amsmath with fixes and additions
\usepackage{booktabs} % commands to create good-looking tables
\usepackage{tikz} % nice language for creating drawings and diagrams
\usetikzlibrary{calc,fit,positioning,arrows.meta,backgrounds,shapes.geometric}

\tikzset{
  vertex/.style={circle, draw, fill=white, inner sep=1.2pt, minimum size=8pt},
  sNode/.style={vertex, fill=green!35},
  tNode/.style={vertex, fill=red!35},
  aNode/.style={vertex, fill=violet!35},
  sphereS/.style={draw=blue, dashed, thick, ellipse, inner xsep=10pt, inner ysep=6pt},
  sphereT/.style={draw=green!50!black, dashed, thick, ellipse, inner xsep=10pt, inner ysep=6pt},
}
\title{SPHERE: Spherical partitioning for large-scale routing optimization}

\author[1]{Robert~Fabian~Lindermann}
\author[1]{Paul\mbox{-}Niklas~Ken~Kandora}
\author[1]{Simon~Caspar~Zeller}
\author[1]{Adrian~Asmund~Fessler}
\author[1]{Steffen~Rebennack}

\affil[1]{%
    Institute for Operations Research\\
    Karlsruhe Institute of Technology\\
    Karlsruhe, Germany\\
    \texttt{\{paul-niklas.kandora, simon.zeller, steffen.rebennack\}@kit.edu}\\
    \texttt{\{robert.lindermann, adrian.fessler\}@student.kit.edu}%
}
  
\begin{document}
  
\maketitle

\begin{abstract}
We study shortest-path routing in large weighted, undirected graphs, where expanding search frontiers raise time and memory costs for exact solvers. We propose \emph{SPHERE}, a query-aware partitioning heuristic that adaptively splits the problem by identifying \emph{source-target} ($s$--$t$) overlaps of hop-distance spheres. Selecting an anchor node $a$ within this overlap partitions the task into independent induced subgraphs for $s\to a$ and $a\to t$, each restricted to its own induced subgraph. If resulting subgraphs remain large, the procedure recurses on that specific subgraph. We provide a formal guarantee that by using the partition cut within the shared overlap, the resulting subpaths preserve feasibility, thereby avoiding the need for boundary repair. Furthermore, \emph{SPHERE} acts as a solver-agnostic framework that naturally exposes parallelism across subproblems. On million-scale road networks, \emph{SPHERE} achieves faster runtimes and smaller optimality gaps than contemporary state-of-the-art partitioning and community-based routing pipelines. Crucially, it also substantially mitigates heavy-tail runtime outliers suffered by standard exact methods, yielding highly stable and predictable execution times across varying queries.
\end{abstract}

\section{Introduction}
Point-to-point routing in large weighted, undirected graphs is a canonical problem in operations research, one that emerged alongside the field’s inception. Exact label-setting or label-correcting methods (e.g., Dijkstra) can certify optimality \citep{Dijkstra1959}, but their search frontiers grow quickly on large networks and become memory- and time-intensive. Graph partitioning methods (e.g., Kernighan-Lin, multilevel partitioning/METIS, spectral cuts) reduce problem size by cutting the graph into parts \citep{KernighanLin1970,KarypisKumar1998,ShiMalik2000}, yet cuts chosen without regard to a specific query $(s,t)$ can intersect narrow passages that many feasible $s$--$t$ routes must use, which can exclude high-quality routes or require costly boundary repair. Bidirectional search reduces depth by growing frontiers from $s$ and $t$ \citep{Nicholson1966,Pohl1969,KaindlKainz1997,Holte2016}, but still explores a single coupled state space and may expand many partial routes on large graphs. For completeness, we note that while feasibility for unconstrained shortest paths is trivial on connected graphs, constrained variants (e.g., resource- or budget-limited paths) are NP-hard; our focus here is the unconstrained case with nonnegative weights, which admits an optimal solution via Dijkstra’s algorithm \citep{Dijkstra1959}.\\
We propose \emph{SPHERE}, a query-aware source-target partitioning heuristic that aligns computation directly with the $s$--$t$ geometry. Let $S_R(s)$ and $S_R(t)$ denote \emph{hop-distance spheres} centered around $s$ and $t$. \emph{SPHERE} initializes a radius tuple $(R_s,R_t)$ large enough to guarantee a nonempty overlap; $S_{R_s}(s) \cap S_{R_t}(t) \neq \emptyset$. \emph{SPHERE} then applies a monotone \emph{decrement} rule to systematically shrink the radii, maintaining a balanced scale ($|R_s - R_t| \le 1$) until reaching the \emph{minimal balanced pair} which refers to the situation where any further valid reduction of either radius would leave the intersection empty. \\
From this minimal overlap, the algorithm selects an anchor node $a \in S_{R_s}(s) \cap S_{R_t}(t)$. This anchor effectively splits the original source-target pair $s$--$t$ into two independent subproblems, $s \to a$ and $a \to t$, which are then solved on their respective induced subgraphs subject to a tunable bound. This tunable bound is a bound that limits the maximum hop-radius of a subproblem which we refer to as $R_{\max}$. If a resulting subproblem still exceeds this cap, the procedure simply recurses on its induced subgraph. \\
Crucially, because the anchor $a$ sits inside the shared overlap, it exists in both induced subgraphs simultaneously. This guarantees that the local $s \to a$ path and the local $a \to t$ path will physically connect at $a$, seamlessly forming a continuous, valid $s$--$t$ route without requiring any secondary algorithms to repair broken links across the cut. Finally, \emph{SPHERE} is solver-agnostic and generates independent subproblems that can be executed in parallel.
We make the following contributions:
\begin{enumerate}
\item \textbf{Query-Aware Geometric Partitioning:} We introduce a heuristic that adaptively partitions the graph based on the specific $s-t$ geometry by cutting inside the last nonempty overlap of hop-distance spheres. This ensures that each part preserves local connectivity and can be rejoined at an anchor node without the need for costly boundary repair.
\item \textbf{Theoretical Guarantees:} We formalize the spherical partitioning algorithm and prove that our cut placement intrinsically preserves global route feasibility. Furthermore, for unweighted, undirected graphs we establish theoretical conditions under which the recursive application of this partition guarantees a globally optimal shortest path.
\item \textbf{Solver-Agnostic Parallel Execution:} The subproblems generated by \emph{SPHERE} are completely independent induced subgraphs. This allows the subproblems to be solved in parallel. Because \emph{SPHERE} generates standard graphs, start nodes, and anchor nodes to the downstream solver, it is completely solver-agnostic. Any off-the-shelf routing algorithm (e.g. Dijkstra) can be used to compute local paths without the need to adjust the solver's underlying structure.
\item \textbf{Empirical validation.} On large-scale networks with over a million nodes and edges, \emph{SPHERE} achieves faster runtimes and smaller optimality gaps than other state-of-the-art $s$--$t$ routing approaches. Crucially, it also substantially mitigates heavy-tail runtime slowdowns observed in standard methods, resulting in highly stable and predictable execution times across varying queries.
\end{enumerate}

\section{Motivation}
Algorithms for routing at scale inherently struggle to balance search space complexity, route exactness, and practical scalability \citep{bast2016route}. Building on these established challenges, routing at scale must meet three criteria simultaneously: control per-instance complexity, guarantee route feasibility, and enable parallel execution with standard solvers. Existing tools typically sacrifice at least one of these:\\
Labeling methods preserve feasibility, but generate a \emph{combinatorial explosion} of partial routes due to the complexity of the topology of large graphs \citep{Feillet2004,IrnichDesaulniers2005}. Graph partitioners such as \citep{KernighanLin1970,KarypisKumar1998,ShiMalik2000} successfully control instance size, but their static boundaries ignore the specific $s$ and $t$ locations. This risks cutting across critical bottlenecks and losing \emph{feasible routes}. Bidirectional search \citep{Nicholson1966,Pohl1969,KaindlKainz1997,Holte2016} reduces depth but explores a single coupled state space that cannot be partitioned into \emph{independent pieces}, making it difficult to enforce parallelism. While static partitioning methods like Corridor routing could theoretically be parallelized, they often require shared boundary information or synchronization across a global quotient graph to ensure path consistency. In contrast, \emph{SPHERE’s} subproblems are mathematically independent by construction. Because the anchor node $a$ is guaranteed to exist in both induced subgraphs, no inter-process communication or boundary repair is required during the solve phase, allowing for true solver-agnostic parallel execution.

In comparison, \emph{SPHERE} avoids \emph{combinatorial explosion} by bounding subproblems within hop-distance spheres, strictly limited by a tunable bound ($R_{\max}$). Additionally, \emph{SPHERE} cuts exclusively inside the shared $s$--$t$ overlap and preserves local connectivity and guarantees a \emph{feasible route} upon concatenation without any need for boundary repair. The induced subgraphs generated by \emph{SPHERE} naturally expose parallel execution and act as a solver-agnostic framework, allowing practitioners to drop-in any trusted downstream optimizer, from exact Mixed-Integer Programming (MIP) to strong metaheuristics \citep{PisingerRopke2007,Vidal2022}, without modification.

In short, \emph{SPHERE} offers a partition strategy that controls complexity, preserves feasibility, and exposes natural parallelism while remaining compatible with the solvers practitioners already trust.

\section{Related Work}
Point-to-point routing on large weighted graphs has been studied from several perspectives. Classical exact methods such as label-setting and label-correcting algorithms (e.g., Dijkstra) can guarantee global optimality on connected graphs with nonnegative weights \citep{Dijkstra1959}. Dynamic-programming and labeling variants extend this to richer settings and can certify optimality, but they often face an explosion in the number of variables on large graphs \citep{Feillet2004,IrnichDesaulniers2005}. Approximation schemes \citep{Hassin1992} exist, trading accuracy for speed, yet these techniques operate on the full graph and therefore do not reduce the instance size before the optimization.

\subsection{Graph Partitioning and Parallel Solvers}
Graph partitioning offers a complementary idea: simplify the problem by splitting the graph, a strategy specifically suited for large-scale routing. Kernighan-Lin improves an initial cut by greedy node swaps to reduce the number of edges crossing between parts while keeping parts balanced \citep{KernighanLin1970}. Multilevel schemes such as METIS repeatedly coarsen the graph, compute a cut on the coarse instance, and refine it during uncoarsening, which gives strong scalability in practice \citep{KarypisKumar1998}. In routing pipelines, these partitions, along with community-based heuristics like the Louvain method, are often used to build a \emph{quotient graph} for a coarse path or corridor that is later refined within parts \citep{Blondel2008}. A limitation of all such methods is that the cut is chosen without regard to the specific start-target pair $(s,t)$, so the boundary can lie across a narrow passage that every feasible $s$--$t$ path must use. Once this happens, the coarse or restricted routing stage may miss or degrade the necessary connection leading to potentially infeasible routes.\\
Parallel computation for shortest-path problems has also been studied to accelerate routing. For instance, \citet{Rozhon2023} analyze parallel Breadth-First Search and approximate distances within subgraphs. However, this approach is mainly theoretically analyzed. 

\subsection{Machine Learning and Deep Learning-Based Approaches}
Recent advancements in neural combinatorial optimization have introduced Deep Learning models capable of solving complex routing tasks. Notable examples include Policy Optimization with Multiple Optima (POMO) \citep{Kwon2020} and Multi-Task Vehicle Routing Solver with Mixture-of-Experts (MvMOE) \citep{Zhou2024}. These neural solvers are primarily designed for distribution-trained variants of the Traveling Salesperson Problem (TSP) and Vehicle Routing Problem (VRP). Those methods operate effectively on small to medium-scale graphs.

\subsection{Bidirectional Search}
Bidirectional search reduces search depth by expanding frontiers from both ends and meeting near the middle \citep{Nicholson1966,Pohl1969,KaindlKainz1997}. The MM algorithm of \citet{Holte2016} provides a priority rule that ensures expansions do not go beyond the midpoint of the optimal solution. These methods improve exploration efficiency but still operate within a single coupled search space and do not produce small, independent instances. In large graphs this can still generate many partial routes, and unlike partition-based methods, they do not offer a straightforward path to parallel, solver-agnostic execution.

\section{Spherical Partitioning}\label{sec:spherical-partitioning}

\subsection{Spherical subgraphs}\label{subsec:spherical-subgraphs}
We begin by formalizing the local "window" around a node using hop-based distance. Throughout this work, let $G=(V,E)$ denote a finite, connected, and undirected graph with positive edge weights. Unless otherwise specified, we measure distance in \emph{hops} (unweighted shortest path), though our formulations remain valid for weighted shortest-path distances. For a comprehensive summary of the mathematical notation used in this paper, please refer to Appendix~\ref{app:notation}.

\begin{definition}[Spheres]\label{def:sph}
For $R\in\mathbb{N}$ and a reference node $\tilde v\in V$, the (closed) \emph{sphere} is
\[
S_R(\tilde v)\;\coloneqq\;\{\,v\in V:\ d_G(\tilde v,v)\le R\,\}\subseteq V,
\]
where $d_G(\tilde v,v)$ is the unweighted shortest path distance between $\tilde v$ and $v$.
\end{definition}
Two standard subgraphs provide complementary local views of $G$ on the node set $S_R(\tilde v)$.

\begin{definition}[Spherical subgraph]\label{def:sph_sub}
The spherical subgraph is defined as $G^{\mathrm{sp}}_R(\tilde v)\coloneqq(V_R(\tilde v),E^{\mathrm{sp}}_R(\tilde v))$
with $V_R(\tilde v) \coloneqq S_R(\tilde v)$ and 
\[
\begin{aligned}
E^{\mathrm{sp}}_R(\tilde v)
\coloneqq
\Bigl\{\{i,j\}\in E :\;&
d_G(\tilde v,i) \le R,\;
d_G(\tilde v,j) \le R,\\
&\bigl|d_G(\tilde v,i)-d_G(\tilde v,j)\bigr| = 1
\Bigr\}.
\end{aligned}
\]

\end{definition}

Thus, $E^{\mathrm{sp}}_R(\tilde v)$ contains exactly those edges inside the sphere that connect consecutive distance layers and therefore lie on shortest paths from $\tilde v$.

\begin{definition}[Induced spherical subgraph]
The induced spherical subgraph, $G^{\mathrm{ind}}_R(\tilde v)\coloneqq(V_R(\tilde v),E^{\mathrm{ind}}_R(\tilde v))$ is defined with
$V_R(\tilde v) \coloneqq S_R(\tilde v)$ and
\[
E^{\mathrm{ind}}_R(\tilde v) \coloneqq \{\{i,j\}\in E:\; i,j\in S_R(\tilde v)\},
\]
where $E^{\mathrm{ind}}_R(\tilde v)$ can be viewed as the edge-closure of the vertex set  $S_R(\tilde v)$, in the sense that all edges of $G$ with both endpoints in $S_R(\tilde v)$ are retained.
\end{definition}

Clearly $E^{\mathrm{sp}}_R(\tilde v)\subseteq E^{\mathrm{ind}}_R(\tilde v)$: the spherical subgraph preserves shortest‑path connectivity to the center, while the induced version preserves all \emph{local} connectivity.

% ----------------- FIG 1: spherical vs induced ------------------------
\begin{figure}[t]
\centering
\tikzset{
  vertex/.style={circle,draw,thick,inner sep=1.5pt,minimum size=8pt},
  sNode/.style={vertex,fill=purple!20,draw=purple!50!black},
  edge/.style={thick},
  sphere/.style={rounded corners=10pt,draw=purple!40,very thick,opacity=.25},
  add/.style={draw=purple!35,thin,dashed}
}
\resizebox{\columnwidth}{!}{%
\begin{tikzpicture}[x=1cm,y=1cm]
% --- (a) spherical subgraph ---
\node at (-0.1,2.0) {\small(a) $G^{\mathrm{sp}}_R(\tilde v)$};
\node[sNode,label=left:$\tilde v$] (s) at (-0.8,0){};
\node[sNode] (u1) at (0.0,1.0) {};
\node[sNode] (u2) at (0.7,0.1) {};
\node[sNode] (u3) at (0.0,-1.0) {};
\node[vertex] (u4) at (1.3,-0.7) {};
\draw[edge, color=purple] (s)--(u1) (s)--(u2) (s)--(u3) ; % shortest-path structure
\draw[edge] (u2)--(u4) (u1)--(u2) (u3)--(u2);
\draw[sphere,fill=purple!15] (-1.4,-1.4) rectangle (1.6,1.4);

% --- (b) induced subgraph ---
\begin{scope}[shift={(4.6,0)}]
\node at (-0.1,2.0) {\small(b) $G^{\mathrm{ind}}_R(\tilde v)$};
\node[sNode,label=left:$\tilde v$] (s2) at (-0.8,0){};
\node[sNode] (v1) at (0.0,1.0) {};
\node[sNode] (v2) at (0.7,0.1) {};
\node[sNode] (v3) at (0.0,-1.0) {};
\node[vertex] (v4) at (1.3,-0.7) {};
\draw[edge, color=purple] (s2)--(v1) (s2)--(v2) (s2)--(v3) (v1)--(v2) (v3)--(v2);
\draw[edge] (v2)--(v4);
%\draw[add] (v1)--(v2) (v1)--(v4) (v3)--(v2); % extra local edges
\draw[sphere,fill=purple!15] (-1.4,-1.4) rectangle (1.6,1.4);
\end{scope}
\end{tikzpicture}}
\caption{\textbf{Two local views on $S_R(\tilde v)$.} Left: spherical subgraph.
Right: induced subgraph.}
\label{fig:sph-vs-induced}
\end{figure}

\subsection{Partition cut}\label{subsec:decomp-cut}
With these geometric building blocks established, we now align these local windows to a specific query $(s,t)$.

\begin{definition}\label{def:overlap}
For radii $(R_s,R_t)\in\mathbb{N}^2$, define the overlap as 
\[
O(R_s,R_t)\;\coloneqq\;S_{R_s}(s)\cap S_{R_t}(t).
\]
\end{definition}

In order to identify a precise splitting point, we control how the radii evolve from their initial state to a minimal configuration. This necessitates defining the properties of \emph{decrement rules}, which systematically shrink the search space.

\begin{definition}[Decrement rule]
\label{def:decru}
A decrement rule
\begin{align*}
\mathrm{decru}:\mathbb{N}^2\to\mathbb{N}^2
\end{align*}

is a map that is monotone in the lexicographic order, i.e.,
\[
\begin{aligned}
(x_1,x_2)\le_{\mathrm{lex}}(y_1,y_2)
\;\Longleftrightarrow\;&
\begin{aligned}[t]
   & x_1 < y_1,\\
   &\text{or }(x_1 = y_1 \text{ and } x_2 \le y_2).
\end{aligned}
\end{aligned}
\]
and never increases any coordinate.
\end{definition}

\begin{example}[monotone decrement]\label{monotonedecrement}
A standard rule that satisfies these properties is one that prioritizes shrinking the larger sphere to keep the search balanced:
\[
\mathrm{decru}(R_s,R_t)=
\begin{cases}
(R_s{-}1,R_t), & R_s\ge R_t,\\[2pt]
(R_s,R_t{-}1), & \text{otherwise.}
\end{cases}
\]
\end{example}
This rule ensures the radii remain balanced ($|R_s - R_t| \leq 1$) throughout the process. If $(R_s,R_t)\le_{\mathrm{lex}}(R'_s,R'_t)$, the smaller pair cannot "overtake" the larger one, preserving the order of operations. 

We make this procedure operational by defining two auxiliary rules for initialization and anchor selection:

\begin{definition}[Starting rule]
\label{def:start}
A starting rule
\begin{align*}
    \mathrm{staru}:V\times V\to\mathbb{N}^2
\end{align*}
%$\mathrm{staru}:V\times V\to\mathbb{N}^2$
returns $(R_s^0,R_t^0)$ with nonempty overlap
$O(R_s^0,R_t^0)\neq\emptyset$ (e.g., $R_s^0{+}R_t^0\ge d_G(s,t)$; a stronger but safe choice is
$S_{R_s^0}(s)\cup S_{R_t^0}(t)=V$).
\end{definition}

\begin{definition}[Anchor rule]
\label{def:anchor}
An anchor rule 
\begin{align*}
    \mathrm{anru}:\mathcal{P}(V)\to V
\end{align*}
%$\mathrm{anru}:\mathcal{P}(V)\to V$
selects $a\in O(R_s,R_t)$
(e.g., uniform random).
\end{definition}
The starting rule for initialization (Definition~\ref{def:start}), decrement rule for shrinking (Definition~\ref{def:decru}) and anchor rule for selecting the split point 
(Definition~\ref{def:anchor}), together define Algorithm~\ref{alg:decomp-cut}, which we refer to as the \emph{Partition cut}.

\begin{algorithm}[ht]
\caption{Partition cut}
\label{alg:decomp-cut}
\begin{algorithmic}[1]
\Require graph $G=(V,E)$, terminals $(s,t)$,
\Statex \hspace{0.97cm} rules $(\mathrm{staru},\mathrm{decru},\mathrm{anru})$
\Ensure \hspace{0.050cm} radii $(\bar R_s,\bar R_t)$, overlap $O$, anchor $a$
\State $(R_s,R_t)\gets \mathrm{staru}(s,t)$
\While{$O(R_s,R_t)\neq\emptyset$}
  \State $(\bar R_s,\bar R_t)\gets (R_s,R_t)$
  \State $(R_s,R_t)\gets \mathrm{decru}(R_s,R_t)$
\EndWhile
\State $O\gets O(\bar R_s,\bar R_t)$,\quad $a\gets \mathrm{anru}(O)$
\State \Return $(\bar R_s,\bar R_t,O,a)$
\end{algorithmic}
\end{algorithm}

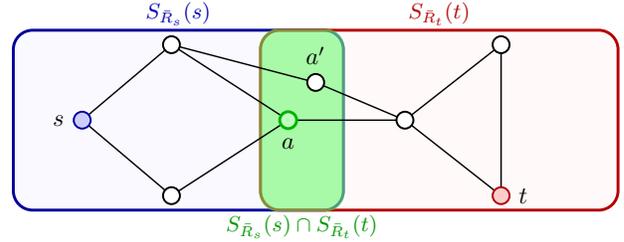
\begin{figure}[ht]
\centering
\def\graphshift{-3mm} % shift ONLY the graph left/right
\tikzset{
  vertex/.style={circle,draw,thick,inner sep=1.2pt,minimum size=7pt,fill=white},
  sNode/.style={vertex,fill=blue!20,draw=blue!60!black},
  tNode/.style={vertex,fill=red!20, draw=red!70!black},
  aNode/.style={vertex,fill=green!25,draw=green!70!black,very thick},
  edge/.style={semithick},
  sphereS/.style={draw=blue!60!black,fill=blue!12,rounded corners=8pt,very thick,fill opacity=.18},
  sphereT/.style={draw=red!70!black, fill=red!12, rounded corners=8pt,very thick,fill opacity=.18}
}
\resizebox{\columnwidth}{!}{%
\begin{tikzpicture}[x=1cm,y=1cm]
% --- fixed anchors for spheres (unshifted) ---
\coordinate (s0)  at (-3.0,  0.0);
\coordinate (u20) at (-1.7,  1.1);
\coordinate (u30) at (-1.7, -1.1);
\coordinate (a0)  at ( 0.00,  0.00);
\coordinate (ap0) at ( 0.40,  0.55);
\coordinate (u50) at ( 1.70,  0.00);
\coordinate (u60) at ( 3.10,  1.10);
\coordinate (t0)  at ( 3.10, -1.10);
\coordinate (tpad0) at (3.80,-1.10); % widen red sphere to cover t

% --- spheres (boxes) at fixed anchors ---
\node[sphereS,fit=(s0) (u20) (u30) (a0) (ap0),
      xshift=-6mm, inner xsep=20pt, inner ysep=6pt] (Sball) {};
\node[sphereT,fit=(t0) (u60) (u50) (a0) (ap0) (tpad0),
      inner xsep=20pt, inner ysep=6pt] (Tball) {};

% --- green intersection (double clip) ---
\begin{scope}
  \clip[rounded corners=8pt] ($(Sball.south west)$) rectangle ($(Sball.north east)$);
  \clip[rounded corners=8pt] ($(Tball.south west)$) rectangle ($(Tball.north east)$);
  \fill[green!60,opacity=.55]
       ($(Sball.south west)-(.3,.3)$) rectangle ($(Tball.north east)+(.3,.3)$);
\end{scope}

% --- shifted graph coordinates ---
\coordinate (shift) at (\graphshift,0);
\foreach \orig/\shft in {s0/s, u20/u2, u30/u3, a0/a, ap0/ap, u50/u5, u60/u6, t0/t} {
  \path (\orig) ++(shift) coordinate (\shft);
}

% --- draw EDGES FIRST (so nodes will cover them) ---
\draw[edge] (s)--(u2) (s)--(u3) (u2)--(a) (u3)--(a)
            (a)--(u5) (u5)--(u6) (u5)--(t) (u6)--(t)
            (u2)--(ap) (ap)--(u5);

% --- draw NODES AFTER edges; nodes are opaque (fill=white/colored) ---
\node[sNode,label=left:$s$]   at (s)  {};
\node[vertex]                  at (u2) {};
\node[vertex]                  at (u3) {};
\node[aNode,label=below:$a$]   at (a)  {};
\node[vertex,label=above:$a'$] at (ap) {};
\node[vertex]                  at (u5) {};
\node[vertex]                  at (u6) {};
\node[tNode,label=right:$t$]   at (t)  {};

% --- overlap formula label BELOW the intersection ---
\coordinate (OverlapBottom) at ($(Sball.south east)!0.5!(Tball.south west)$);
\node[text=green!60!black]
  at ($(OverlapBottom)+(0,-0.22)$) {\small $S_{\bar R_s}(s)\cap S_{\bar R_t}(t)$};

% --- sphere labels ---
\node[blue!70!black] at ($(Sball.north)+(0,0.22)$) {\small $S_{\bar R_s}(s)$};
\node[red!70!black]  at ($(Tball.north)+(0,0.22)$) {\small $S_{\bar R_t}(t)$}; % moved up
\end{tikzpicture}}
\caption{\textbf{Last nonempty overlap.} Blue and red spheres overlap in the green region containing both $a$ and $a'$. Potential anchor node $a$ is selected according to Definition~\ref{def:anchor}.}
\label{fig:last-overlap}
\end{figure}

The core intuition behind the Partition Cut (visualized in Figure~\ref{fig:last-overlap}) is to identify the smallest possible geometric region where the influence of $s$ and $t$ meet. Initially, spheres are grown around the terminals until they intersect. We then apply the decrement rule, $\mathrm{decru}$, to systematically shrink these radii until any further reduction would result in an empty intersection.

By selecting an anchor node $a$ within this final, minimal overlap (Definition~\ref{def:anchor}), we effectively decompose the original $s \to t$ query into two independent $s \to a$ and $a \to t$ subproblems. These subproblems are then solved within their respective induced spherical subgraphs.

% ----------------- FIG 2: last overlap + anchor (expanded/shaded) ------

A natural question is whether the Partition cut of Algorithm~\ref{alg:decomp-cut} actually preserves feasibility of the original $s{\to}t$ routing problem. In other words: \emph{Does splicing the two subpaths at the chosen anchor still yield a valid global route?} The following statement makes this precise:

\begin{proposition}[Feasibility preservation]\label{prop:feas}
Let $(\bar R_s,\bar R_t,O,a)$ be returned by Algorithm~\ref{alg:decomp-cut}.
If $P_{s\to a}$ is a feasible path in $G^{\mathrm{ind}}_{\bar R_s}(s)$ and
$P_{a\to t}$ is a feasible path in $G^{\mathrm{ind}}_{\bar R_t}(t)$,
then $P_{s\to a}\!\circ\!P_{a\to t}$ is a feasible $s\to t$ path in $G$.
\end{proposition}

\begin{proof}
We provide a short proof sketch: Both subgraphs are induced, hence every used edge is in $G$. Since $a\in O(\bar R_s,\bar R_t)$, concatenating at $a$ yields a valid $s\to t$ path in $G$.
\end{proof}

\subsection{Recursive spherical partitioning}\label{subsec:recursive}
The following procedure extends the partition cut of Algorithm~\ref{alg:decomp-cut} into a recursive framework. The objective is to decompose a global $s \to t$ routing task into a collection of manageable subproblems, each strictly confined to an induced spherical subgraph where the radius does not exceed a bound $R_{\max}$. For a comprehensive overview of all configurable parameters, please refer to Appendix~\ref{app:params}.

Each subproblem is formalized as a \textit{task triple} $(H, u, w)$, representing a localized search from node $u$ to node $w$ within the induced subgraph $H$. We treat the resulting task set $\mathcal{T}$ as a sequence ordered along the $s \to t$ trajectory. This ordering ensures that concatenating the locally optimal paths induces a valid global path. For simplicity, we refer to this locally optimal path as the path returned by Algorithm~\ref{alg:sph-partition}.

\begin{algorithm}[ht]
\caption{Recursive spherical partitioning}
\label{alg:sph-partition}

% ---- compact local setup (only for this algorithm) ----
\begingroup
\makeatletter
\algrenewcommand\algorithmicindent{0.75em}            % tighter nesting
\algrenewcommand\alglinenumber[1]{\scriptsize #1}      % smaller line numbers
\algrenewcommand\ALG@beginalgorithmic{%
  \small                                               % smaller font in the block
  \setlength{\ALG@thistlm}{0pt}% remove extra left inner margin
  \setlength{\abovedisplayskip}{4pt}% tighter display math spacing
  \setlength{\belowdisplayskip}{4pt}%
}
\makeatother
% ------------------------------------------------------

\begin{algorithmic}[1]
\Require graph $G=(V,E)$, terminals $(s,t)$,
\Statex \hspace{0.86cm}  bound $R_{\max}$, rules $\mathrm{staru},\mathrm{decru},\mathrm{anru}$% as in Alg.~\ref{alg:decomp-cut}
\Ensure \hspace{0.05cm} set $\mathcal{T}$ of task triples $(H,u,w)$
\Function{SphPartition}{$G,s,t,R_{\max}$}
  \State $\mathcal{T}\gets\emptyset$, $\texttt{rule}\leftarrow (\mathrm{staru},\mathrm{decru},\mathrm{anru}) $
  \State $(R_s,R_t,O,a)\gets \Call{PartCut}{G,s,t,\texttt{rule}}$ \hfill \(\triangleright\)\,Alg.~\ref{alg:decomp-cut}
  \If{$R_s\le R_{\max}$}
     \State $\mathcal{T}\gets \mathcal{T}\cup\{(G^{\mathrm{ind}}_{R_s}(s),\,s,\,a)\}$
  \Else
     \State $\mathcal{T}\gets \mathcal{T}\cup \Call{SphPartition}{G^{\mathrm{ind}}_{R_s}(s),\,s,\,a,\,R_{\max}}$
  \EndIf
  \If{$R_t\le R_{\max}$}
     \State $\mathcal{T}\gets \mathcal{T}\cup\{(G^{\mathrm{ind}}_{R_t}(t),\,a,\,t)\}$
  \Else
     \State $\mathcal{T}\gets \mathcal{T}\cup \Call{SphPartition}{G^{\mathrm{ind}}_{R_t}(t),\,a,\,t,\,R_{\max}}$
  \EndIf
  \State \Return $\mathcal{T}$
\EndFunction
\end{algorithmic}
\endgroup
\end{algorithm}
Algorithm~\ref{alg:sph-partition} starts by initializing the set of tasks $\mathcal{T}$ to be empty (lines~1-3). It then performs a partition cut on the original problem $(G,s,t)$, which returns the radii $(R_s,R_t)$, the last nonempty overlap $O$, and an anchor $a\in O$. The anchor acts as a splitting point that separates the global $s\!\to\!t$ route into two sides. \\
The algorithm then branches into two recursive checks. On the $s$-side (lines~4-7), the relevant subproblem is $(G^{\mathrm{ind}}_{R_s}(s),s,a)$. If the radius $R_s$ is already below the cap 
$R_{\max}$, this subproblem is considered sufficiently small and is added to the task set $\mathcal{T}$. If not, the procedure calls itself recursively on this induced subgraph to split it further. Symmetrically, the $t$-side branch (lines~8-11) processes the subproblem $(G^{\mathrm{ind}}_{R_t}(t),a,t)$: either it is recorded directly if $R_t\le R_{\max}$, or it is further partitioned by another recursive call. \\
Finally, once both branches have been processed, the algorithm returns (line~12) the accumulated 
set $\mathcal{T}$ of task triples. By construction, concatenating the solutions to these recorded subproblems along their anchors yields a valid overall $s\!\to\!t$ route.

% ----------------- FIG 3: recursion schematic (unchanged) --------------
\begin{figure}[ht]
\centering
\tikzset{
  box/.style={rounded corners,draw=black!50,thick,fill=black!3,inner sep=2.5pt},
  lab/.style={font=\scriptsize},
  arr/.style={-{Stealth[length=2.2mm]},semithick}
}
\resizebox{\columnwidth}{!}{%
\begin{tikzpicture}[x=1cm,y=1cm]
\node[box,align=center] (root) {$(G,s,t)$\\ \scriptsize Partition cut $\Rightarrow$ $(R_s,R_t,O,a)$};
\node[box,below left=8mm and 7mm of root] (L) {$G^{\mathrm{ind}}_{R_s}(s),\ s\to a$};
\node[box,below right=8mm and 7mm of root] (R) {$G^{\mathrm{ind}}_{R_t}(t),\ a\to t$};
\draw[arr] (root.south west) -- (L.north);
\draw[arr] (root.south east) -- (R.north);
\node[lab,below=1.5mm of L] (Lbl) {if $R_s\le R_{\max}$: record; else: recurse};
\node[lab,below=1.5mm of R] (Rbl) {if $R_t\le R_{\max}$: record; else: recurse};
\end{tikzpicture}}
\caption{\textbf{Recursive splitting.} Any side above the bound is split again by
Algorithm~\ref{alg:decomp-cut}; leaves are independent subproblems.}
\label{fig:recursion-schematic}
\end{figure}
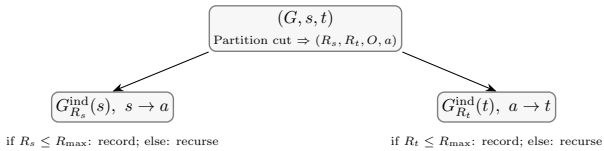
Figure~\ref{fig:recursion-schematic} illustrates this recursive structure. Each partition cut splits the problem into two boxes, one for the start-side and one for the target-side. If the radius on a side is within the bound, the box is recorded as a leaf; otherwise, it is expanded by another partition cut.

\subsubsection{Baseline application}
The recursive decomposition defined in Algorithm~\ref{alg:sph-partition} yields a set of leaf tasks, each formalized as $(H, u, w)$. This process effectively reduces the high-complexity global $s \to t$ routing problem into a collection of independent, localized subproblems. Because these subproblems are spatially confined and bounded by the bound $R_{\max}$, they are sufficiently small to be solved efficiently using standard routing methods or off-the-shelf optimizers.

\begin{algorithm}[ht]
\caption{Baseline application with spherical partitioning}
\label{alg:baseline-application}
\begin{algorithmic}[1]
\Require graph $G=(V,E)$, terminals $(s,t)$, 
\Statex \hspace{0.97cm} bound $R_{\max}$, solver $\mathrm{optimize}(\cdot)$
\Ensure feasible route $\mathcal{R}$
\State $\mathcal{R} \gets \emptyset$; \quad $\mathcal{T} \gets \Call{SphPartition}{G, s, t, R_{\max}}$ 
\For{\textbf{each} $(H, u, w) \in \mathcal{T}$} 
    \State $\mathcal{R} \gets \mathcal{R} \circ \mathrm{optimize}(H, u, w)$
\EndFor
\State \Return $\mathcal{R}$ 
\end{algorithmic}
\end{algorithm}

Algorithm~\ref{alg:baseline-application} details the integration of recursive spherical partitioning into a complete routing pipeline. The procedure begins by decomposing the input graph $G$ and terminals $(s,t)$ into a sequence of smaller, manageable subproblems with resulting task $(H,u,w)$.

Subsequently, an arbitrary downstream solver, $\mathrm{optimize}(H,u,w)$, is applied to each subproblem to compute a local path. These local solutions are then concatenated in the order established by the partition's anchors, forming a globally valid $s \to t$ route $\mathcal{R}$. A critical advantage of this framework is that because the subproblems are mutually independent, they can be solved in parallel. Furthermore, the procedure remains strictly solver-agnostic, allowing the utilization of various downstream optimizer without modification to the underlying partitioning logic.

\subsection{Optimality Conditions}
We now consider the special case of a finite, connected, undirected, and \emph{unweighted} graph $G=(V,E)$. As before, let $d(\cdot,\cdot)$ denote the unweighted shortest-path distance, or \emph{hop count}. Building upon the definition of the sphere $S_R(\tilde{v})$ for a node $\tilde{v} \in V$ and radius $R \in \mathbb{N}$ (Definition~\ref{def:sph}) , we revisit the overlap $O(R_s, R_t)$ established in Definition~\ref{def:overlap} for a given pair of radii $(R_s, R_t)$.

In the following, we demonstrate that Algorithm~\ref{alg:decomp-cut}, when applied to unweighted graphs using the monotone decrement rule from Example~\ref{monotonedecrement}, preserves optimality. We thus provide a stronger result compared to the the general weighted setting, where Proposition~\ref{prop:feas} ensures the preservation of route feasibility only.

Furthermore, this result allows to compute a globally shortest path without invoking any additional optimization routines or solver-specific procedures, provided we apply the recursion until the radii are equal to $1$. 

We begin by formalizing the \emph{Strong Minimality Condition}, which characterizes the minimal non-empty overlap induced by the monotone decrement rule in terms of the associated radius tuple $(R_s,R_t)$.
\begin{definition}[Strong Minimality Condition]\label{def:strong-min}
A tuple $(R_s,R_t)$ is \emph{strongly minimal} if
\[
\begin{aligned}
& O(R_s,R_t)\neq\emptyset, \\
& O(R_s-1,R_t)=\emptyset, \\
& O(R_s,R_t-1)=\emptyset.
\end{aligned}
\]
\end{definition}

The next result shows by combining Definitions~\ref{def:overlap} and~\ref{def:strong-min} that strong minimality forces all overlap vertices to lie simultaneously on the boundaries of both spheres determined by $(R_s,R_t)$.

\begin{lemma}[Boundary property]\label{lem:boundary}
If $(R_s,R_t)$ is strongly minimal, then for every $v\in O(R_s,R_t)$,
\[
d(s,v)=R_s
\quad\text{and}\quad
d(t,v)=R_t.
\]
\end{lemma}

\begin{proof}
Let $v\in O(R_s,R_t)$. If $d(s,v)\le R_s-1$, then $v\in O(R_s-1,R_t)$,
contradicting strong minimality. Hence $d(s,v)=R_s$. The argumentation for $t$
is analogous, yielding $d(t,v)=R_t$.
\end{proof}

Having established the boundary property of strongly minimal overlaps,
we now derive the key consequence of the monotone decrement rule.
The following \emph{Meet-in-the-middle equality} formalizes that a strongly minimal radius tuple satisfies
\[
R_s + R_t = d(s,t),
\]
thereby identifying the last overlap of two spheres.

\begin{lemma}[Meet-in-the-middle equality]\label{lem:mitm}
If $(R_s,R_t)$ is strongly minimal, then
\[
R_s+R_t = d(s,t).
\]
\end{lemma}

\begin{proof}
Choose $a\in O(R_s,R_t)$. Since $a\in S_{R_s}(s)$ and $a\in S_{R_t}(t)$,
there exists an $s$--$t$ path via $a$ of length at most $R_s+R_t$, hence
$d(s,t)\le R_s+R_t$.\\
Assume for contradiction that $d(s,t)\le R_s+R_t-1$. Let $P$ be a shortest
$s$--$t$ path of length $L=d(s,t)$. Let $v$ be the vertex on $P$ at distance $R_s$ from $s$. Since $P$ is the shortest $s$--$t$ path, then $d(s,v)=R_s$ and $d(t,v)=L-R_s\le R_t-1$. Thus $v\in O(R_s,R_t-1)$ contradicts strong
minimality. Therefore $d(s,t)\ge R_s+R_t$, and equality follows.
\end{proof}

Lemma~\ref{lem:mitm} constitutes the central argument of our main optimality theorem: it ensures that any anchor selected from the minimal overlap lies on a globally shortest path and thus enables recursive preservation of optimality. Note that all definitions and lemmas apply to any terminal pair $(u,w)$ (replacing $(s,t)$), in every recursive subproblem.

% \begin{theorem}[Global optimality under recursive partitioning with $R_{\max}=1$]\label{thm:global-opt-rmax1}
% Consider a recursive procedure that, for any subproblem with endpoints
% $(u,w)$, returns a strongly minimal pair $(R_u,R_w)$ (via a sequential decrement
% rule) and selects an anchor $a\in O(R_u,R_w)$, then recurses on $(u,a)$ and
% $(a,w)$ until all leaves satisfy $R_{\max}=1$.

% Then the final $s$--$t$ path obtained by concatenating the leaf subpaths is a
% globally shortest path in $G$.
% \end{theorem}

\begin{theorem}[Global preservation]\label{thm:global-opt-rmax1}
Let $G=(V,E)$ be a finite, connected, undirected, and unweighted graph and assume that the monotone decrement rule is used. %For every subproblem with endpoints $(u,w)$, 
%Algorithm \ref{alg:decomp-cut}:
%\begin{enumerate}
%\item computes a strongly minimal radius tuple $(R_u,R_w)$ via the sequential monotone decrement rule,
%\item selects an anchor $a \in O(R_u,R_w)$, and
%\item the procedure is recursively applied to $(u,a)$ and $(a,w)$.
%\end{enumerate}
Then the concatenated $s$--$t$ path returned by Algorithm~\ref{alg:sph-partition} is a globally shortest path in $G$.
\end{theorem}

\begin{proof}
We argue by induction over the recursion tree:

\emph{Local optimal split.}
Fix any subproblem with endpoints $(u,w)$ and a strongly minimal pair
$(R_u,R_w)$ as induced by undirected graphs and the monotone decrement rule with anchor $a\in O(R_u,R_w)$. Applying Lemma~\ref{lem:boundary}, we observe that the anchor $a$ lies on the boundaries of both spheres, such that $d(u,a)=R_u$ and $d(a,w)=R_w$. It follows from Lemma~\ref{lem:mitm} that these radii sum exactly to the distance between $u$ and $w$, i.e. $R_u+R_w=d(u,w)$.
Hence
\[
d(u,a)+d(a,w)=d(u,w),
\]
so $a$ lies on a shortest $u$--$w$ path and the split $(u,w)\mapsto (u,a),(a,w)$
is consistent with global optimality for that subproblem.

\emph{Inductive step.}
Assume recursively computed solutions are shortest for the two child subproblems
$(u,a)$ and $(a,w)$. Their concatenation has length
$d(u,a)+d(a,w)=d(u,w)$, hence is shortest for the parent subproblem.
Applying this from leaves to the root yields a globally shortest $s$--$t$ path.
\end{proof}

\section{Numerical Experiments}\label{sec:numerical_experiments}
Our experiments are designed to evaluate whether our method delivers significant performance gains (in terms of runtime and optimality gap) over established graph-routing baselines.

\subsection{Dataset}
For our empirical evaluation, we use the \texttt{West-USA-road} benchmark suite from the 9th DIMACS Implementation Challenge on Shortest Paths \citep{DIMACS2006SP}. This dataset represents real-world road networks. Nodes correspond to intersections and edges represent road segments with positive weights equal to physical length. 

\subsection{Experimental Setup}
For each problem instance \(p\) (with according instance seed), we draw a source-target pair \((s,t)\) uniformly at random. To capture variability from the randomized anchor selection, we run five independent trials with inner seeds \(q\) for each source-target pair. We report results aggregated by the median and mean across these seeds. Further details regarding graph dimensions and seed selection are provided in Appendix~\ref{app:implementation}.

We apply Algorithm~\ref{alg:sph-partition} with a maximum radius bound of $R_{\max}=1800$ (see Appendix~\ref{app:implementation} for tuning rationale). Because \emph{SPHERE} generates independent induced subgraphs, we solve the resulting leaf tasks in parallel using Dijkstra's algorithm and report the end-to-end wall-clock time, including the partitioning and concatenation overhead.\footnote{The graph data and implementation will be made publicly available upon the manuscript’s acceptance.} 

%\paragraph{Baselines.} 
We compare \emph{SPHERE} against an optimal reference using Dijkstra's algorithm on the full graph, as well as Louvain and Corridor routing. Unlike \emph{SPHERE}, these static partitioning schemes compute cuts globally rather than conditioning on the specific $(s,t)$ query. Consequently, they cannot guarantee feasibility without boundary repair, nor do they naturally yield the mathematically independent subproblems required for the same drop-in parallel execution framework applied to our method. While static partitions could theoretically be parallelized, they often require shared boundary synchronization or global quotient graphs to ensure path consistency, whereas \emph{SPHERE} subproblems are self-contained by construction. (See Appendix~\ref{app:implementation} for implementation and library details).

%\noindent\textbf{Metrics.} 
Let \(m \in \{\emph{SPHERE},\ \textsc{Corridor},\ \textsc{Louvain}\}\) denote the method. For each problem seed \(p\) and inner seed \(q\), let \(C_{p,q}^{(m)}\) be the route cost returned by \(m\), and let \(C_p^\star\) be the shortest-path cost from Dijkstra’s algorithm on the full graph for the same \((s,t)\). Define the relative optimality gap
\[
\Delta_{p,q}^{(m)} \;=\; \frac{C_{p,q}^{(m)} - C_p^\star}{C_p^\star}.
\]
For each \(p\), aggregate across the five inner seeds by the median:
\begin{align*}
\widetilde{\Delta}_{p}^{(m)} \;=\; \operatorname{median}_{q=1,\dots,5}\,\Delta_{p,q}^{(m)},\\
\widetilde{T}_{p}^{(m)} \;=\; \operatorname{median}_{q=1,\dots,5}\,T_{p,q}^{(m)},
\end{align*}
where \(T_{p,q}^{(m)}\) is the end-to-end wall-clock time. For completeness, we also report means:
\[
\overline{\Delta}_{p}^{(m)} \;=\; \tfrac{1}{5}\sum_{q=1}^{5}\Delta_{p,q}^{(m)},\qquad
\overline{T}_{p}^{(m)} \;=\; \tfrac{1}{5}\sum_{q=1}^{5}T_{p,q}^{(m)}.
\]

\subsection{Results}\label{sec:results}
Across 30 West-USA instances with five inner seeds each, \emph{SPHERE} achieves faster end-to-end runtimes than the three baselines and matches near-optimal route quality. We use performance and accuracy profiles for visualization \citep{dolan2002benchmarking,beiranvand2017best}. Both are computed per instance from the median over seeds. Mean-based profiles for runtime and accuracy are presented as well.

\begin{figure*}[t]
  \centering
  % top row
  \subcaptionbox{Runtime profile based on the \textbf{median} over five seeds
    \label{fig:profile_runtime_median}}{%
    \includegraphics{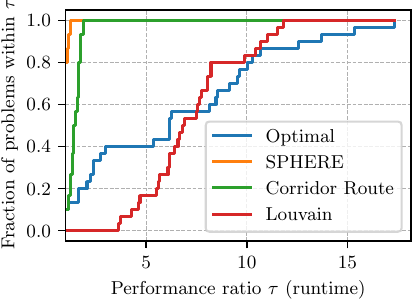}}
  \hfill
  \subcaptionbox{Accuracy profile based on the \textbf{median} over five seeds
    \label{fig:profile_accuracy_median}}{%
    \includegraphics{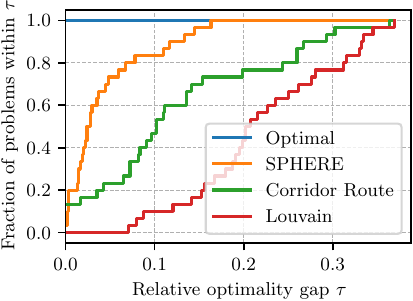}}

  \par\medskip

  % bottom row
  \subcaptionbox{Runtime profile based on the \textbf{mean} over five seeds
    \label{fig:profile_runtime_mean}}{%
    \includegraphics{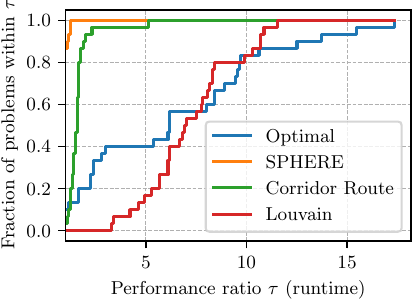}}
  \hfill
  \subcaptionbox{Accuracy profile based on the \textbf{mean} over five seeds
    \label{fig:profile_accuracy_mean}}{%
    \includegraphics{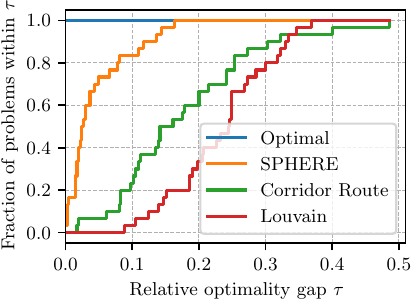}}

  \caption{Performance and accuracy profiles of 30 instances based on median and mean over five seeds.}
  \label{fig:profiles_westusa}
\end{figure*}

%\paragraph{Runtime.}
Figures~\ref{fig:profile_runtime_median} and \ref{fig:profile_runtime_mean} plot the fraction of instances solved within a performance ratio~$\tau$ of the best method on that instance. \emph{SPHERE’s} curve lies above the Louvain and corridor routes for most $\tau$, indicating uniformly stronger or equal runtime across instances. Interestingly, for small $\tau$, Louvain’s curve lies below Dijkstra’s, despite being a suboptimal method. However, at larger $\tau$ it crosses and stays above, indicating tail slowdowns for Dijkstra. This suggests that \emph{SPHERE} not only improves median runtime, but also avoids the heavy-tailed slowdowns observed in both Dijkstra and Louvain, resulting in tighter runtime distributions and fewer extreme outliers, which makes \emph{SPHERE’s} runtimes more predictable across different instances.

%\paragraph{Route quality.}
Figures~\ref{fig:profile_accuracy_median} and \ref{fig:profile_accuracy_mean} report the fraction of instances whose relative optimality gap is at most~$\tau$. Dijkstra attains zero gap by construction. \emph{SPHERE’s} curve dominates the partition baselines across~$\tau$, yielding smaller or equal gaps on the majority of instances. This confirms that cutting inside the $s$--$t$ overlap and solving on induced spheres preserves high-quality routes without boundary repair. 

% \paragraph{Pareto dominance}
Against the partition baselines (Louvain and METIS corridor), \emph{SPHERE} Pareto-dominates on 26 of 30 instances in terms of average metrics. It is both faster and has a smaller optimality gap (Appendix~\ref{app:perf_per_instance}). The remaining three show a speed-quality trade-off where \emph{SPHERE} is not dominated by another algorithm. In other words, \emph{SPHERE} consistently lies on or close to the empirical speed–quality frontier, while Louvain and corridor-routing often face a trade-off between being faster but less accurate, or more accurate but substantially slower.

%\paragraph{Robustness and aggregation}
Repeating the analysis with means over seeds yields profiles that do not differ materially from the median-based plots, supporting robustness to the randomized anchor selection. Additionally, the full results in Appendix~\ref{app:perf_per_instance} show low dispersion across seeds. The results indicate reliability across varying query geometries and suggest that spherical decomposition effectively captures the core topological features necessary for high-quality routing, regardless of internal stochasticity. For \emph{SPHERE}, the optimality-gap standard deviation is 0 on 16 instances and is the lowest among all methods on all but one instance. For further information, we refer to Appendix~\ref{app:perf_per_instance}.

\section{Conclusion and Outlook}
We introduced \emph{SPHERE}, a routing scheme that splits an $s$--$t$ query by cutting \emph{inside} the last nonempty overlap of hop-spheres around $s$ and $t$. This placement preserves feasibility without boundary repair, confines each side to an induced subgraph, and allows the subproblems to be solved independently with standard shortest-path routines and in parallel. On million-scale road networks, \emph{SPHERE} achieves faster runtimes and smaller optimality gaps than state-of-the-art static and community-based partitioning methods. Notably, our approach eliminates the heavy-tail runtime outliers common in exact solvers, providing high statistical stability across randomized anchor selections.

Future work will explore data-driven anchor selection to further improve robustness across diverse graph topologies. Additionally, \emph{SPHERE} can serve as a wrapper for complex neural combinatorial optimizers that typically struggle with large-scale instances. Another direction could be to automate the tuning of the radius cap $R_{\max}$ to adaptively balance computational budgets against query complexity. These advancements will maintain \emph{SPHERE}'s simplicity and parallelism while further enhancing speed and stability.
\newpage
\bibliography{references}

\appendix
\section{Appendix}
\subsection{Notation}\label{app:notation}
\begin{longtblr}[
  caption = {Notation}, label = {tab:notation}
]{colspec = {@{} Q[l,wd=4.6cm] X @{}}, rowhead = 2}
\toprule
Symbol & Description \\
\midrule
$V=\{1,\dots,m\}$ & Node (vertex) index set. \\
$\mathcal{P}(V)$ & Power set of $V$. \\
$E\subseteq V\times V$ & Edge set; undirected unless stated. \\
$G=(V,E)$ & Graph with node set $V$ and edge set $E$. \\
$d_G(u,v)$ & Shortest unweighted path distance on $G$. \\
$G[U]$ & Induced subgraph on node set $U$. \\
$S_R(\tilde v)=\{\,v\in V:\ d_G(\tilde v,v)\le R\,\}$ & (Closed) hop sphere (ball) of radius $R$ around $\tilde v$. \\
$S_s(r)\equiv S_r(s),\; S_t(r)\equiv S_r(t)$ & Shorthand spheres around $s$ and $t$. \\
$O(R_s,R_t)=S_{R_s}(s)\cap S_{R_t}(t)$ & Overlap of the two spheres. \\
$a\in V$,\; $S^\dagger\subseteq O(R_s,R_t)$ & Anchor node; optional candidate anchor set. \\
$G_s=G[S_{R_s}(s)],\; G_t=G[S_{R_t}(t)]$ & Induced subgraphs around $s$ and $t$ at radii $R_s,R_t$. \\
$(H,u,w)$,\; $\mathcal{T}$ & Task triple “route $u\to w$ inside $H$”; set of leaf tasks. \\
$|X|$ & Cardinality of a (finite) set $X$. \\
\bottomrule
\end{longtblr}

\subsection{Parameters}\label{app:params}
\begin{longtblr}[
  caption = {Parameters}, label = {tab:params}
]{colspec = {@{} Q[l,wd=3.4cm] X @{}}, rowhead = 2}
\toprule
Symbol & Description \\
\midrule
$R_{\max}\in\mathbb{N}$ & Maximum allowed radius for spherical subgraphs (controls subproblem diameter). \\
$V_{\max}\in\mathbb{N}$,\; $E_{\max}\in\mathbb{N}$ & Optional caps on nodes/edges permitted in any subgraph $G[U]$. \\
$\varepsilon_{\text{overlap}}\in\mathbb{N}_0$ & Minimum required overlap size (require $|O(R_s,R_t)|\ge\varepsilon_{\text{overlap}}$). \\
$\Delta r\in\mathbb{N}$ & Radius increment/decrement step during sphere growth/shrinking (if using fixed steps; typically $\Delta r=1$). \\
$L_{\max}\in\mathbb{N}$ & Maximum recursion depth for repeated splits. \\
$k_{\text{anchor}}\in\mathbb{N}$ & Number of anchors to select from $O$ (often $k_{\text{anchor}}=1$). \\
\bottomrule
\end{longtblr}

\subsection{Implementation Details and Hyperparameters}\label{app:implementation}
\paragraph{Graph Dimensions.}
The \texttt{West-USA-road} network consists of $6,262,104$ nodes and $15,248,146$ edges. The large scale of this network makes it a challenging benchmark for standard exact solvers.

\paragraph{Instance and Seed Selection.}
The $30$ problem instances $p$ are defined by fixed global seeds to ensure reproducibility of the $(s,t)$ pairs. For each $p$, $5$ inner seeds $q$ control the stochasticity of the $\mathrm{anru}$ function in Algorithm~\ref{alg:decomp-cut}.

\paragraph{Radius bound ($R_{\max}$).} 
The tunable bound $R_{\max}$ controls the maximum diameter of the subproblems generated by Algorithm~\ref{alg:baseline-application}. Throughout development, we established a robust rule of thumb: when routing within a spherical subgraph of hop radius $R$, choosing $R_{\max} \in [\,R/2,\; R/1.7\,]$ delivers high-quality solutions at both low computational cost and effort. For the West-USA benchmark instances, a single sphere with $R \approx 3200$ generally covers the necessary region. Thus, our chosen bound of $R_{\max} = 1800$ lies squarely in this optimal band, carefully balancing the likelihood of an $s$--$t$ overlap against the complexity of the resulting leaf tasks.

\paragraph{Baseline Configurations.}
The baseline pipelines were implemented using standard libraries to ensure a fair comparison. The corridor-routing baseline utilizes \texttt{PyMetis} to compute the initial $k=64$ cell partition. The Louvain community-detection baseline was run using its default modularity-optimization parameters, seeded where applicable to ensure deterministic execution. It is worth noting that because these static partitioning schemes compute cuts globally rather than conditioning on the specific $(s,t)$ query, they cannot guarantee feasibility without boundary repair, nor do they yield strictly independent leaves suitable for the same parallel execution framework applied to \emph{SPHERE}.
\subsection{Performance per instance}
\label{app:perf_per_instance}
Continued on next page.
\onecolumn
\begin{longtable}{@{} l l c c c c @{}}
\caption{WestUSA -- Routing}\label{tab:westusa-routing}\\
\toprule
& & \textbf{avg compute time} & \textbf{avg opt. gap} & \textbf{median opt. gap} & \textbf{std opt. gap} \\
\midrule
\endfirsthead
\toprule
& & \textbf{avg compute time} & \textbf{avg opt. gap} & \textbf{median opt. gap} & \textbf{std opt. gap} \\
\midrule
\endhead
\midrule
\multicolumn{6}{r}{Continued on next page}\\
\endfoot
\bottomrule
\endlastfoot
\multirow{3}{*}{\textbf{P1}} & \emph{SPHERE} & \textbf{135.73} & \textbf{0.08} & \textbf{0.06} & \textbf{0.02} \\
 & Corridor Route & 173.97 & 0.24 & 0.07 & 0.36 \\
 & Louvain & 1047.77 & 0.25 & 0.28 & 0.05 \\
\addlinespace
\multirow{3}{*}{\textbf{P2}} & \emph{SPHERE} & \textbf{138.27} & \textbf{0.12} & 0.12 & \textbf{0.00} \\
 & Corridor Route & 187.87 & 0.14 & \textbf{0.09} & 0.16 \\
 & Louvain & 1072.17 & 0.34 & 0.35 & 0.05 \\
\addlinespace
\multirow{3}{*}{\textbf{P3}} & \emph{SPHERE} & \textbf{165.25} & \textbf{0.03} & \textbf{0.03} & \textbf{0.00} \\
 & Corridor Route & 190.12 & 0.08 & 0.10 & 0.06 \\
 & Louvain & 1130.18 & 0.18 & 0.18 & 0.05 \\
\addlinespace
\multirow{3}{*}{\textbf{P4}} & \emph{SPHERE} & \textbf{370.06} & \textbf{0.11} & \textbf{0.11} & \textbf{0.00} \\
 & Corridor Route & 574.18 & 0.25 & 0.26 & 0.10 \\
 & Louvain & 1230.71 & 0.37 & 0.37 & 0.03 \\
\addlinespace
\multirow{3}{*}{\textbf{P5}} & \emph{SPHERE} & \textbf{126.88} & \textbf{0.05} & \textbf{0.05} & \textbf{0.00} \\
 & Corridor Route & 171.35 & 0.10 & 0.11 & 0.06 \\
 & Louvain & 1129.89 & 0.14 & 0.14 & 0.06 \\
\addlinespace
\multirow{3}{*}{\textbf{P6}} & \emph{SPHERE} & \textbf{301.07} & \textbf{0.02} & \textbf{0.03} & \textbf{0.00} \\
 & Corridor Route & 571.56 & 0.18 & 0.15 & 0.14 \\
 & Louvain & 1265.49 & 0.25 & 0.20 & 0.12 \\
\addlinespace
\multirow{3}{*}{\textbf{P7}} & \emph{SPHERE} & \textbf{213.59} & \textbf{0.00} & \textbf{0.00} & \textbf{0.00} \\
 & Corridor Route & 354.04 & 0.27 & 0.36 & 0.17 \\
 & Louvain & 1197.91 & 0.32 & 0.32 & 0.04 \\
\addlinespace
\multirow{3}{*}{\textbf{P8}} & \emph{SPHERE} & \textbf{157.02} & \textbf{0.01} & \textbf{0.01} & \textbf{0.00} \\
 & Corridor Route & 224.89 & 0.32 & 0.27 & 0.16 \\
 & Louvain & 1165.53 & 0.09 & 0.07 & 0.03 \\
\addlinespace
\multirow{3}{*}{\textbf{P9}} & \emph{SPHERE} & \textbf{390.68} & \textbf{0.02} & \textbf{0.02} & \textbf{0.00} \\
 & Corridor Route & 396.99 & 0.13 & 0.10 & 0.10 \\
 & Louvain & 1274.68 & 0.23 & 0.21 & 0.06 \\
\addlinespace
\multirow{3}{*}{\textbf{P10}} & \emph{SPHERE} & \textbf{137.20} & 0.02 & 0.02 & \textbf{0.00} \\
 & Corridor Route & 197.31 & \textbf{0.02} & \textbf{0.00} & 0.03 \\
 & Louvain & 1141.72 & 0.27 & 0.33 & 0.10 \\
\addlinespace
\multirow{3}{*}{\textbf{P11}} & \emph{SPHERE} & \textbf{130.96} & \textbf{0.13} & 0.13 & 0.01 \\
 & Corridor Route & 167.76 & 0.20 & \textbf{0.07} & 0.24 \\
 & Louvain & 1057.36 & 0.33 & 0.33 & \textbf{0.00} \\
\addlinespace
\multirow{3}{*}{\textbf{P12}} & \emph{SPHERE} & 264.38 & \textbf{0.04} & \textbf{0.04} & \textbf{0.00} \\
 & Corridor Route & \textbf{234.53} & 0.24 & 0.20 & 0.13 \\
 & Louvain & 1160.17 & 0.25 & 0.17 & 0.13 \\
\addlinespace
\multirow{3}{*}{\textbf{P13}} & \emph{SPHERE} & \textbf{123.22} & \textbf{0.00} & 0.00 & \textbf{0.00} \\
 & Corridor Route & 200.92 & 0.10 & \textbf{0.00} & 0.16 \\
 & Louvain & 1308.92 & 0.18 & 0.19 & 0.03 \\
\addlinespace
\multirow{3}{*}{\textbf{P14}} & \emph{SPHERE} & \textbf{234.34} & \textbf{0.14} & \textbf{0.14} & \textbf{0.00} \\
 & Corridor Route & 358.43 & 0.25 & 0.24 & 0.17 \\
 & Louvain & 1414.78 & 0.33 & 0.31 & 0.05 \\
\addlinespace
\multirow{3}{*}{\textbf{P15}} & \emph{SPHERE} & \textbf{211.37} & \textbf{0.00} & \textbf{0.00} & \textbf{0.00} \\
 & Corridor Route & 300.88 & 0.11 & 0.10 & 0.09 \\
 & Louvain & 1288.33 & 0.15 & 0.15 & 0.02 \\
\addlinespace
\multirow{3}{*}{\textbf{P16}} & \emph{SPHERE} & \textbf{136.40} & \textbf{0.04} & 0.04 & \textbf{0.00} \\
 & Corridor Route & 221.84 & 0.08 & \textbf{0.02} & 0.13 \\
 & Louvain & 1400.16 & 0.20 & 0.19 & 0.05 \\
\addlinespace
\multirow{3}{*}{\textbf{P17}} & \emph{SPHERE} & \textbf{119.90} & \textbf{0.01} & 0.00 & \textbf{0.02} \\
 & Corridor Route & 193.90 & 0.02 & \textbf{0.00} & 0.03 \\
 & Louvain & 1266.58 & 0.20 & 0.21 & 0.07 \\
\addlinespace
\multirow{3}{*}{\textbf{P18}} & \emph{SPHERE} & \textbf{123.18} & 0.08 & 0.08 & \textbf{0.01} \\
 & Corridor Route & 205.16 & \textbf{0.06} & \textbf{0.00} & 0.07 \\
 & Louvain & 1409.96 & 0.32 & 0.33 & 0.03 \\
\addlinespace
\multirow{3}{*}{\textbf{P19}} & \emph{SPHERE} & \textbf{205.97} & \textbf{0.02} & \textbf{0.02} & \textbf{0.00} \\
 & Corridor Route & 1052.15 & 0.40 & 0.30 & 0.31 \\
 & Louvain & 1401.55 & 0.25 & 0.25 & 0.05 \\
\addlinespace
\multirow{3}{*}{\textbf{P20}} & \emph{SPHERE} & \textbf{170.12} & \textbf{0.00} & \textbf{0.00} & \textbf{0.00} \\
 & Corridor Route & 391.16 & 0.30 & 0.08 & 0.34 \\
 & Louvain & 1414.17 & 0.10 & 0.08 & 0.04 \\
\addlinespace
\multirow{3}{*}{\textbf{P21}} & \emph{SPHERE} & \textbf{326.25} & \textbf{0.01} & \textbf{0.01} & \textbf{0.00} \\
 & Corridor Route & 506.65 & 0.20 & 0.14 & 0.12 \\
 & Louvain & 1502.53 & 0.30 & 0.22 & 0.16 \\
\addlinespace
\multirow{3}{*}{\textbf{P22}} & \emph{SPHERE} & \textbf{137.27} & \textbf{0.03} & \textbf{0.03} & \textbf{0.00} \\
 & Corridor Route & 180.60 & 0.49 & 0.29 & 0.51 \\
 & Louvain & 1135.72 & 0.28 & 0.26 & 0.08 \\
\addlinespace
\multirow{3}{*}{\textbf{P23}} & \emph{SPHERE} & \textbf{172.50} & \textbf{0.01} & \textbf{0.01} & \textbf{0.00} \\
 & Corridor Route & 211.28 & 0.08 & 0.04 & 0.11 \\
 & Louvain & 1065.85 & 0.21 & 0.20 & 0.06 \\
\addlinespace
\multirow{3}{*}{\textbf{P24}} & \emph{SPHERE} & \textbf{145.29} & \textbf{0.02} & \textbf{0.02} & \textbf{0.00} \\
 & Corridor Route & 202.11 & 0.21 & 0.26 & 0.13 \\
 & Louvain & 1009.77 & 0.27 & 0.28 & 0.07 \\
\addlinespace
\multirow{3}{*}{\textbf{P25}} & \emph{SPHERE} & \textbf{160.44} & \textbf{0.02} & \textbf{0.02} & \textbf{0.00} \\
 & Corridor Route & 219.89 & 0.11 & 0.04 & 0.14 \\
 & Louvain & 1059.09 & 0.12 & 0.12 & 0.02 \\
\addlinespace
\multirow{3}{*}{\textbf{P26}} & \emph{SPHERE} & \textbf{172.71} & \textbf{0.03} & \textbf{0.03} & \textbf{0.00} \\
 & Corridor Route & 203.42 & 0.17 & 0.13 & 0.14 \\
 & Louvain & 1054.63 & 0.24 & 0.19 & 0.13 \\
\addlinespace
\multirow{3}{*}{\textbf{P27}} & \emph{SPHERE} & \textbf{246.94} & \textbf{0.02} & \textbf{0.02} & \textbf{0.00} \\
 & Corridor Route & 479.93 & 0.14 & 0.06 & 0.11 \\
 & Louvain & 1388.28 & 0.22 & 0.20 & 0.04 \\
\addlinespace
\multirow{3}{*}{\textbf{P28}} & \emph{SPHERE} & \textbf{159.42} & \textbf{0.00} & \textbf{0.00} & \textbf{0.00} \\
 & Corridor Route & 267.58 & 0.16 & 0.13 & 0.17 \\
 & Louvain & 1274.86 & 0.19 & 0.16 & 0.08 \\
\addlinespace
\\
\multirow{3}{*}{\textbf{P29}} & \emph{SPHERE} & \textbf{232.46} & \textbf{0.07} & \textbf{0.07} & \textbf{0.00} \\
 & Corridor Route & 354.39 & 0.14 & 0.11 & 0.07 \\
 & Louvain & 1219.90 & 0.25 & 0.24 & 0.03 \\
\addlinespace
\multirow{3}{*}{\textbf{P30}} & \emph{SPHERE} & \textbf{142.64} & 0.16 & 0.16 & \textbf{0.00} \\
 & Corridor Route & 175.93 & \textbf{0.10} & \textbf{0.08} & 0.10 \\
 & Louvain & 1162.54 & 0.15 & 0.09 & 0.10 \\
\end{longtable}

\twocolumn

\end{document}